\documentclass[twocolumn,showpacs,prl]{revtex4}

\usepackage{graphicx}

\begin{document}

\def\bn{\begin{eqnarray}}
\def\en{\end{eqnarray}}
\def\va{{\vec{a}}}
\def\vb{{\vec{b}}}
\def\vc{{\vec{c}}}
\def\vq{{\vec{q}}}
\def\vr{{\vec{r}}}
\def\vx{{\vec{x}}}
\def\vy{{\vec{y}}}
\def\vrp{{\vec{r}'}}
\def\vN{{\vec{\nabla}}}
\def\bh{{\bf h}}
\def\bl{{\bf l}}
\def\bN{{\bf\nabla}}
\def\bA{{\bf A}}
\def\bM{{\bf M}}
\def\bS{{\bf S}}
\def\bU{{\bf U}}
\def\bW{{\bf W}}
\def\bV{{\bf V}}
\def\d{\delta}
\def\f{\frac}
\def\p{\partial}
\def\s{\sigma}
\def\v{\varepsilon}
\def\L{\Lambda}
\def\lg{\langle}
\def\rg{\rangle}
\def\tB{{\tilde{B}}}
\def\tJ{{\tilde{J}}}
\def\th{{\tilde{h}}}
\def\tp{{\tilde{\psi}}}
\def\qcr{{\vq\cdot\vr}}
\def\qacr{{\vq^\ast\cdot\vr}}

\title{Understanding visual map formation through vortex dynamics \\ of spin
  Hamiltonian models}
\author{Myoung Won Cho}
\email{mwcho@postech.edu}
\author{Seunghwan Kim}
\email{swan@postech.edu}
\affiliation{
  Asia Pacific Center for Theoretical Physics $\&$ NCSL,
  Department of Physics, Pohang University of Science and Technology,
  Kyungpook, Pohang, 790-784, South Korea}
\date{\today}
\begin{abstract}
%We introduce a general method for cerebral cortical map generation and apply
% it to pattern formation in orientation and ocular dominance columns of the brain.
The pattern formation in orientation and ocular dominance columns is one of the
most investigated problems in the brain.
From a known cortical structure, we build spin-like Hamiltonian models with
long-range interactions of the Mexican hat type.
These Hamiltonian models allow a coherent interpretation of the diverse
phenomena in the visual map formation with the help of relaxation dynamics of
spin systems.
In particular, we explain various phenomena of self-organization in orientation
and ocular dominance map formation including the pinwheel annihilation and its
dependency on the columnar wave vector and boundary conditions.
\end{abstract}
\maketitle

A series of experiments have suggested that important elements of the
organization of ocular dominance (OD) and orientation preference (OP) maps in
the striate cortex are not prespecified but emergent during an activity-driven,
self-organizing process~\cite{Hubel1977,LeVay1978,Stryker1978}.
An optical imaging technique~\cite{Blasdel1992,Blasdel1986,Grinvald1986}
revealed the detailed maps of OD and OP over small patches of the cortex,
which prompted several models for the map generation in the brain and various
attempts for the analysis of the properties in the observed cortical map
patterns~\cite{Erwin1995,Swindale1996}. \par

In the experimentally observed OP columnar patterns, there are two prominent
features: (1) singular points (so called ``pinwheels'') are point-like
discontinuities around which the orientation preference changes by multiples of
180$^\circ$ along a closed loop, and (2) linear zones are regions where
iso-orientation contours (IOCs) are straight and run in parallel for a
considerable distance~\cite{Swindale1987,Bonhoeffer1991,Maldonado1997}.
The analysis of competitive Hebbian models~\cite{Obermayer1992,Durbin1990}
predicted the bifurcation between homogeneous and inhomogeneous solutions
depending on the cooperation range $\sigma$ and the change in the wavelength
$\L$~\cite{Hoffsummer1995,Scherf1999,Goodhill2000}.
Linear zones in OP columns or OD bands are features of the inhomogeneous states.
Some experimental or simulational results suggested that pinwheels are not
permanent structures but can be annihilated during the course of
active-dependent development~\cite{Wolf1998,Daw1998}.
The perpendicular intersection of IOCs and OD bands with the margin of the
striate cortex has been also reported~\cite{Obermayer1993,LeVay1985}.
There are some evidence that patterns in OP and OD columns are not independent
but correlated.
Pinwheels have a tendency to align with the centers of OD bands and IOCs
intersect the borders of OD bands at a steep angle~\cite{Obermayer1993}.
The influence of the interactions between OP and OD columns on the pinwheel
stability was described~\cite{Wolf1998}. \par

In this paper, we propose a new visual map formation method with neighborhood
interactions.
The cortical map generation is described by spin-like Hamiltonians with
distance dependent interactions.
The statistical analysis of these Hamiltonians leads to the successful
description of several pattern properties observed {\em in vivo}.
In our analogy, the progress in the visual map formation corresponds to
relaxation dynamics in spin systems, where, for example, the pinwheels
in OP can be regarded as (in-plane) vortices.
The pinwheel instability and its annihilation rate are explained in terms of
the free energy of the topological excitation or the Kosterlitz and Thouless
transition temperature~\cite{Kosterlitz1973}.
Our model with neighborhood interaction exhibits the bifurcation between the
homogeneous and inhomogeneous states depending on the inhibitory activity
strength $k$ in lateral currents.
The columnar wavelength and the correlation function of the OP maps are also
derived from out model.
The extension of our model to the $O(3)$ or the Heisenberg model induces the
correlation between the OP and OD columns, which leads to the orthogonality
between IOCs and the borders of OD bands, and allows OD columns to influence
on the pinwheel stability.
Another orthogonal tendency of the cortical maps with area boundaries is
explained from the equilibrium condition. \par

The six layers in the neocortex can be classified into three different
functional types.
The layer IV neurons first get the long range input currents, such as signals
from the retina via the lateral geniculate nucleus (LGN) of the thalamus, and
send them up vertically to layer II and III that are called the true
association cortex.
Output signals are sent down to the layer V and VI, and sent further to the
thalamus or other deep and distant neural structures.
Lateral connections also occur in the superficial (layer II and III) pyramidal
neurons and have usual antagonistic propensity which sharpens responsiveness to
an area.
However, the superficial pyramidal neurons also send excitatory recurrent to
adjacent neurons due to unmyelinated collaterals.
Horizontal or lateral connections have such distance dependent (so called
"Mexican hat" shaped) excitatory or inhibitory activity.
Some bunches of neurons make columnar clusters called minicolumns and such
aggregations are needed to consider higher dimensional properties of processing
elements~\cite{Calvin1998,Lund1998}. \par

Now we propose a Lie algebraic representation method for the feature vectors in
cortical maps, called the {\it fibre bundle map} (FBM) model.
This starts from the assumption that the total space $E$ is composed of the
lattice (or base) space $B$ and the pattern (or fibre) space $F$.
A transition function (or symmetry) group $G$ of a homeomorphism of the fibre
space $F$ is necessary to describe what happens if there is ``excitatory'' or
``inhibitory'' activity.
The symmetry group $G$ is important since it determines what interactions are
possible.
In the visual cortex, $G$ takes $O(3)$ rather than $O(2)\times Z_2$ because of
the correlations between OP and OD columns. \par

If there are stimuli from lateral and LGN neurons in the OP columns, the
changes in preferred angles $\phi_i$ ($0\le\phi_i<\pi)$ for $i$-th neuron group
are described by
\bn \f{\p \phi_i}{\p t}
%=-\f{\p H}{\p \phi_i}
  &=&-2\v\sum_jI(\vr_i,\vr_j)\sin(2\phi_i-2\phi_j) \nonumber \\
  &&-2\mu B_i\sin(2\phi_i-2\phi_i'), \label{eq:Dynamics} \en
%  =-\v\sum_jI(\vr_i,\vr_j)\sin(2\phi_i-2\phi_j)
%  -\mu B_i\sin(2\phi_i-2\phi_i'), \label{eq:Dynamics} \en
where $\v$ and $\mu$ are the change rates in the stimuli from lateral and LGN
cells, respectively, and $B$ and $\phi'$ are the strength and the phase of LGN
stimulus.
We use the lateral neighborhood interaction function $I$, modified from a
wavelet, as
\bn I_{WL}(\vr_i,\vr_j)=\left(1-k\f{|\vr_i-\vr_j|^2}{\s^2}\right)
  \exp(-\f{|\vr_i-\vr_j|^2}{2\s^2}). \en
Note that there is also another well-known Mexican hat shaped function called
the difference of Gaussians (DOG) filter,
$I_{DOG}(r)=\exp(-r^2/2\s_1^2)-k\exp(-r^2/2\s_2^2)$.
We rewrite Eq.~(\ref{eq:Dynamics}) as a gradient flow with the
Hamiltonian
\bn H&=&-\sum_{i,j}J(\vr_i,\vr_j)\bS_i\cdot\bS_j-\sum_i \bh_i\cdot \bS_i,
  \label{eq:Hamiltonian}
\en
where $J(\vr_i,\vr_j)=\f{\v}{2}I(\vr_i,\vr_j)$ is site distance dependent
interaction energy.
The site states $\bS_i=(\cos 2\phi_i,\sin 2\phi_i)$ and the external stimuli
$\bh_i=(\mu B_i\cos 2\phi_i', \mu B_i\sin 2\phi_i')$ are 2-components vectors.
In this form, $H$ is similar to spin models in statistical physics.
In the case of OD maps, $\bS_i$ has a one-component with $\pm 1$ values similar
to the Ising model.
There were some previous studies on spin models with distance dependent
interactions, albeit mostly in the form of $J(r)\sim r^{-\alpha}$
~\cite{Monroe1990,Luijten1997,Krech2000,Ifti2001}. \par

\begin{figure}[t] 
\rotatebox{90}{ \begin{minipage}[b]{4.5cm} $\tJ(0)-\tJ(q)$ \end{minipage} }
\begin{minipage}[b]{6cm} \includegraphics[width=6cm]{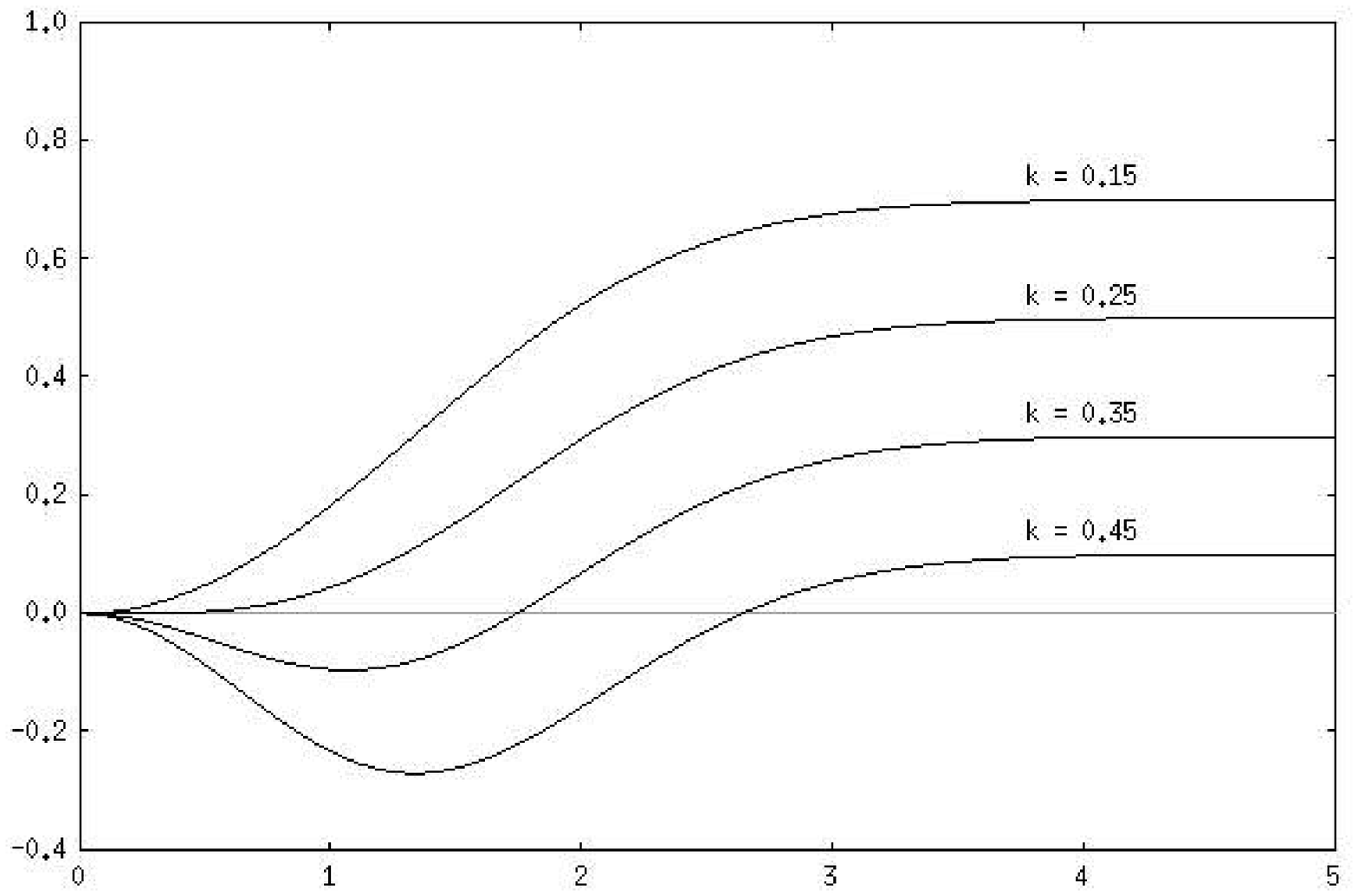} $q$ \end{minipage}
\caption{\label{fig:Jq}
  The plot of $\tJ(0)-\tJ(q)$ vs $q$ in Eq.~(\ref{eq:Jq}).
  For $k>k_c$ (=$1/4$), the minimum point at $q^\ast=0$ becomes unstable.}
\end{figure}

\begin{figure}[t]
\begin{minipage}[b]{2cm} \ \end{minipage}
\begin{minipage}[b]{2cm} \centering $t = 5$ \\ \end{minipage}
\begin{minipage}[b]{2cm} \centering $t = 15$ \\ \end{minipage}
\begin{minipage}[b]{2cm} \centering $t = 100$ \\ \end{minipage} \\ \ \\
\begin{minipage}[b]{2cm} \centering (a) $k = 0.2$ \ \\ \ \\ \ \\ \end{minipage}
\begin{minipage}[b]{2cm} \includegraphics[width=1.9cm]{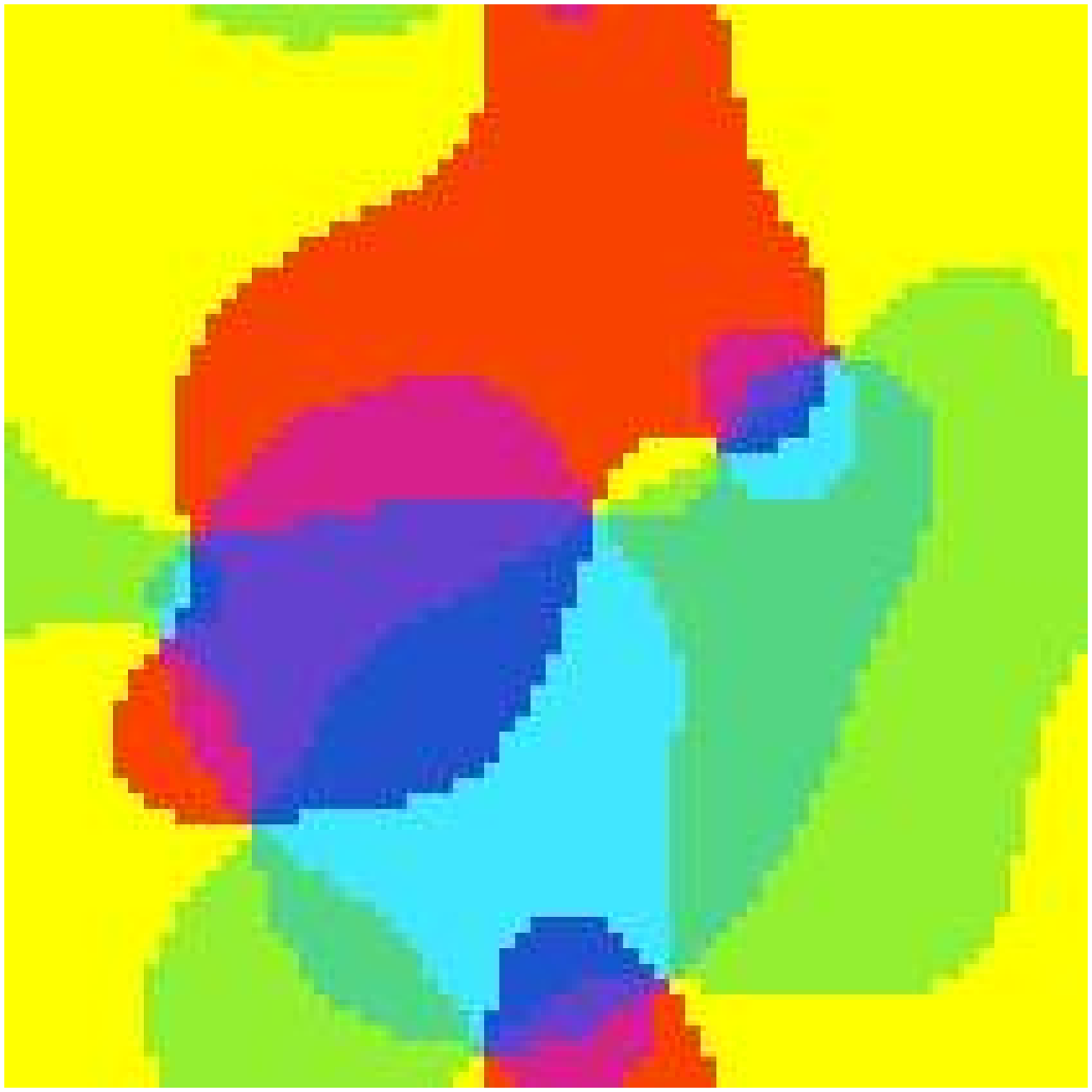} \end{minipage}
\begin{minipage}[b]{2cm} \includegraphics[width=1.9cm]{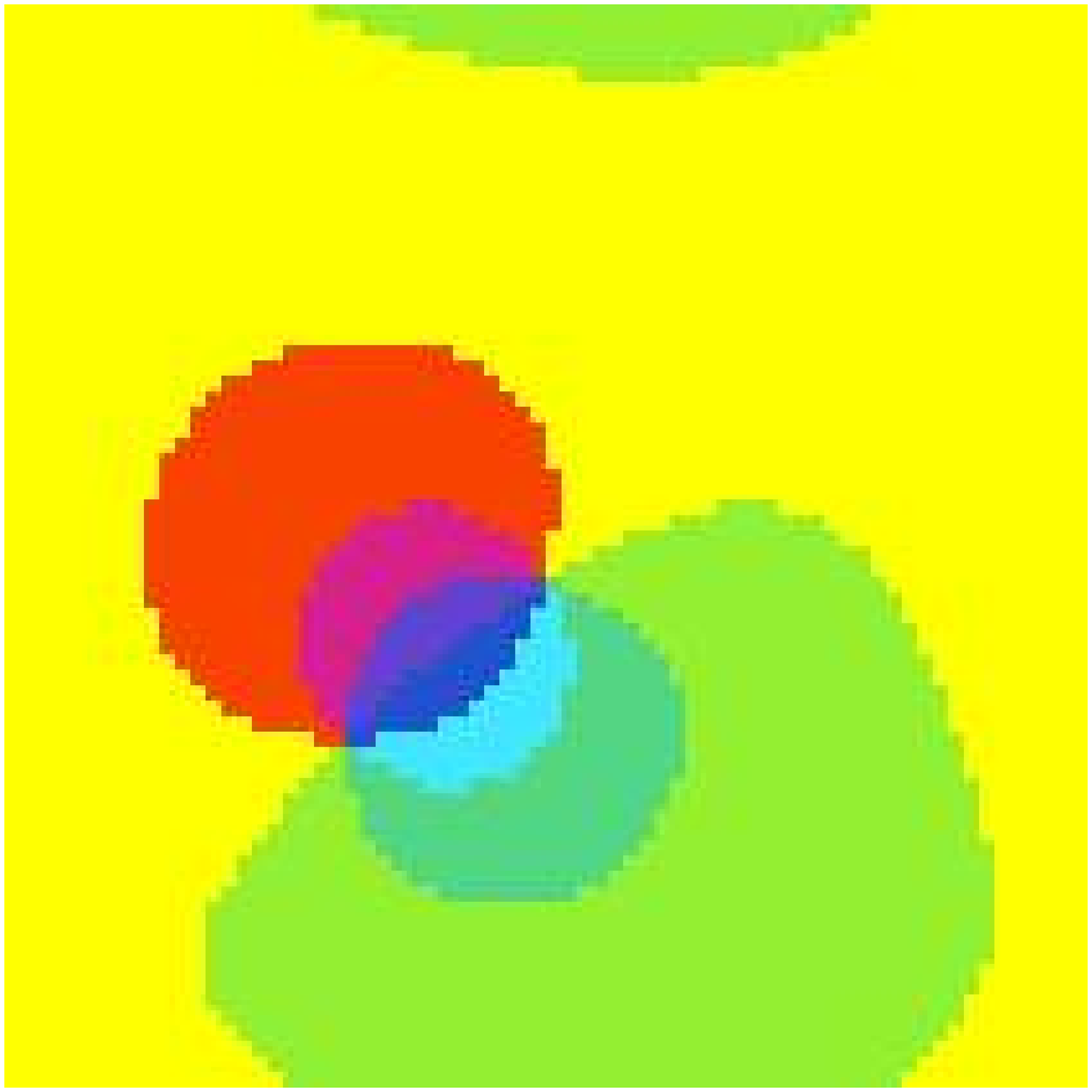} \end{minipage}
\begin{minipage}[b]{2cm} \includegraphics[width=1.9cm]{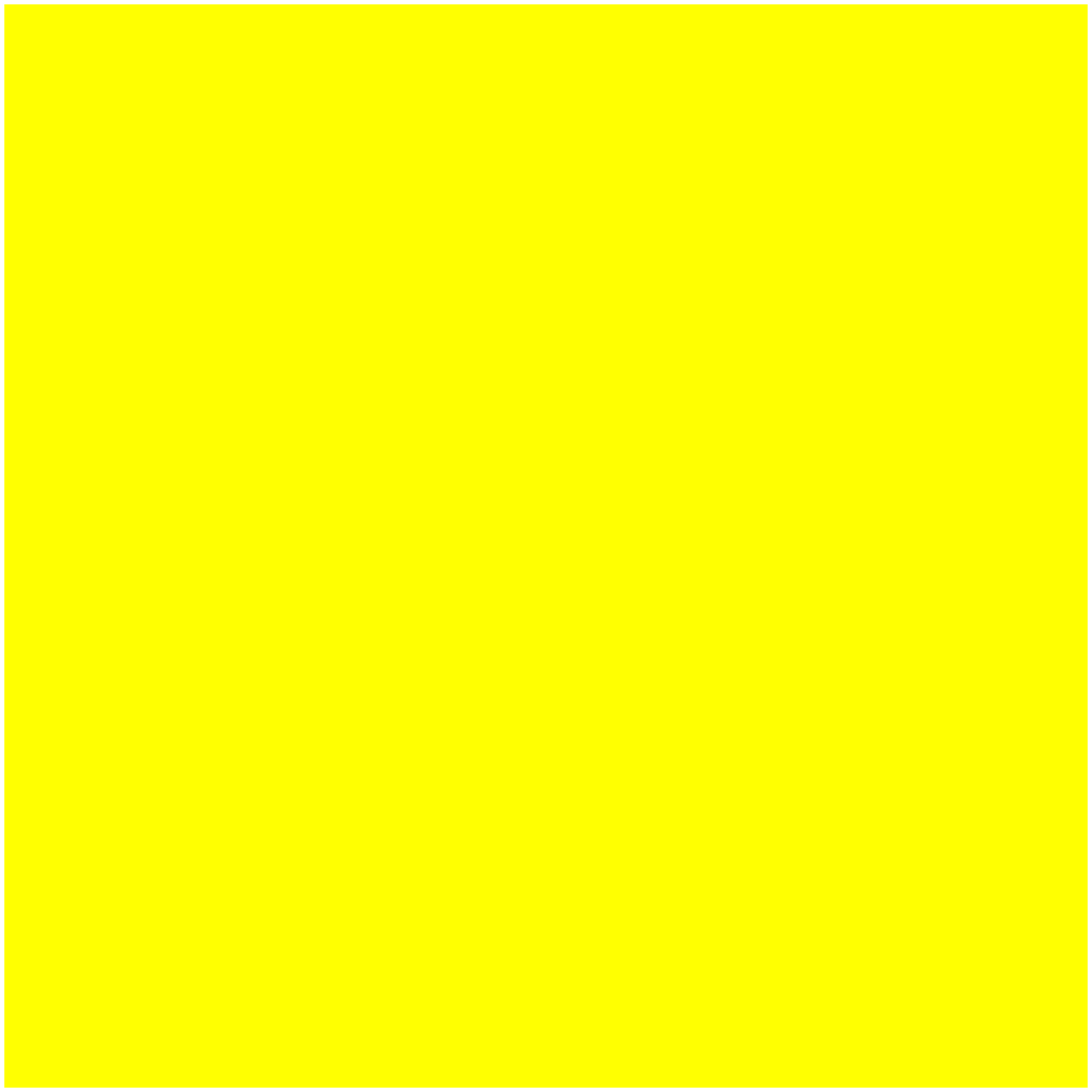} \end{minipage}
\\ \ \\
\begin{minipage}[b]{2cm} \centering (b) $k = 0.3$ \ \\ \ \\ \ \\ \end{minipage}
\begin{minipage}[b]{2cm} \includegraphics[width=1.9cm]{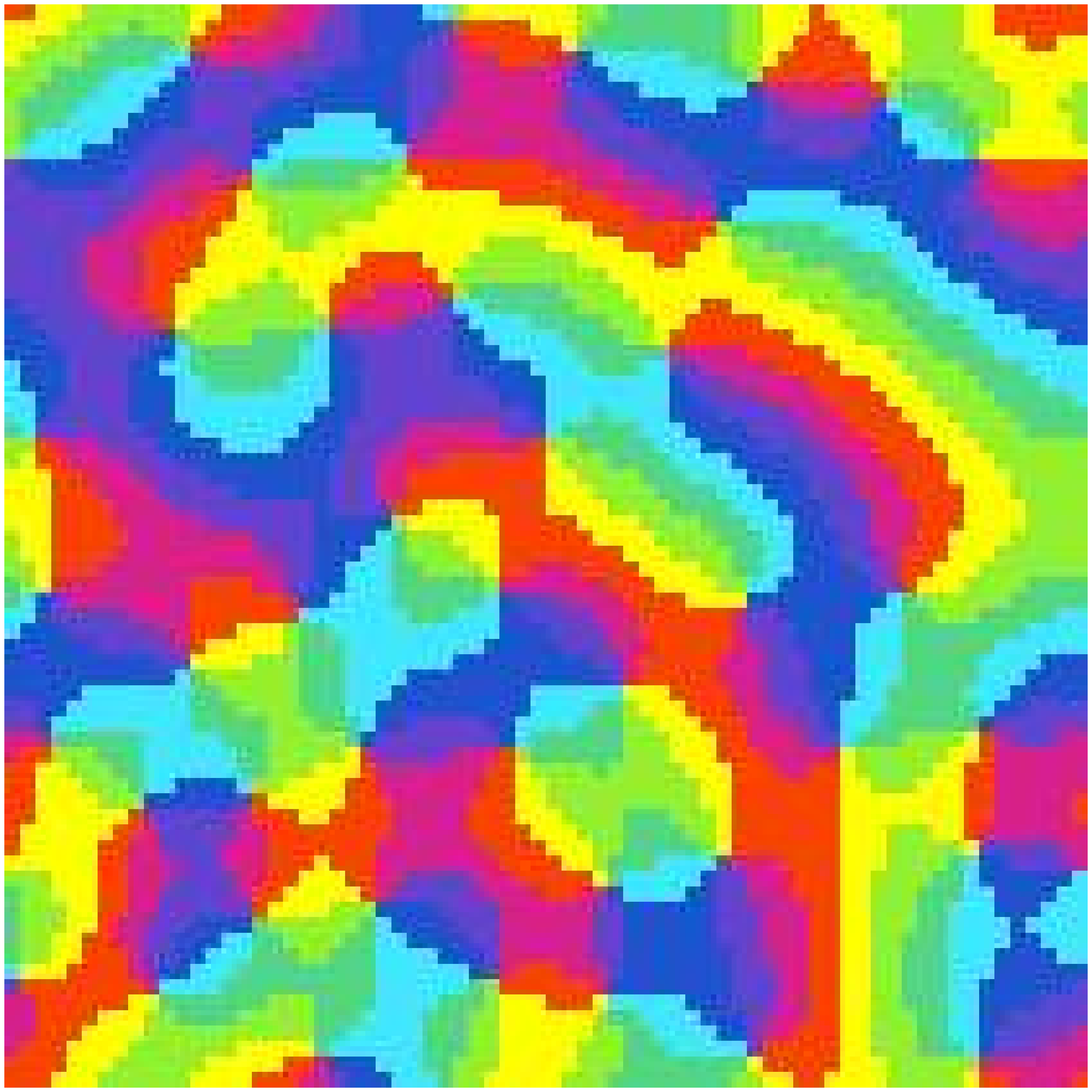} \end{minipage}
\begin{minipage}[b]{2cm} \includegraphics[width=1.9cm]{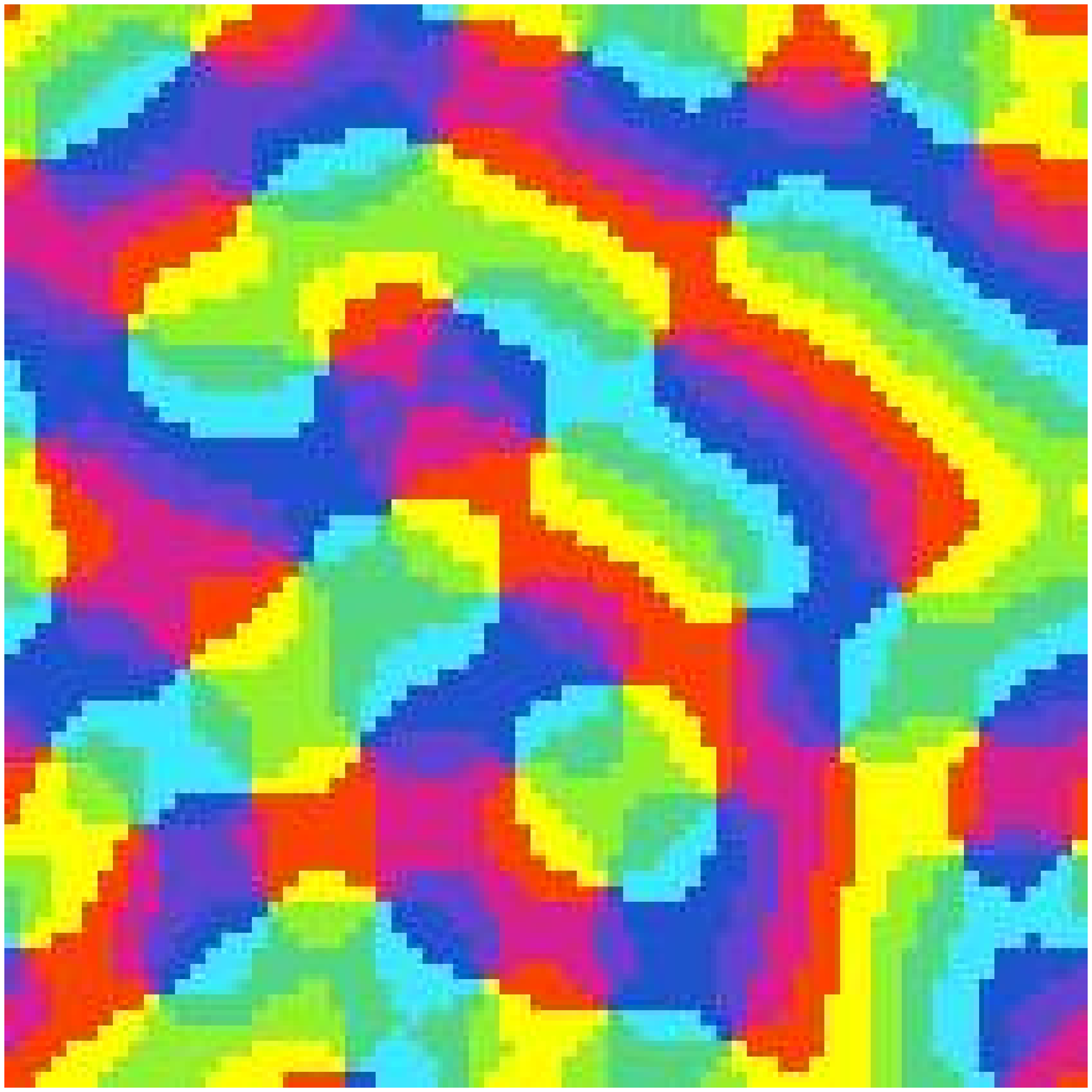} \end{minipage}
\begin{minipage}[b]{2cm} \includegraphics[width=1.9cm]{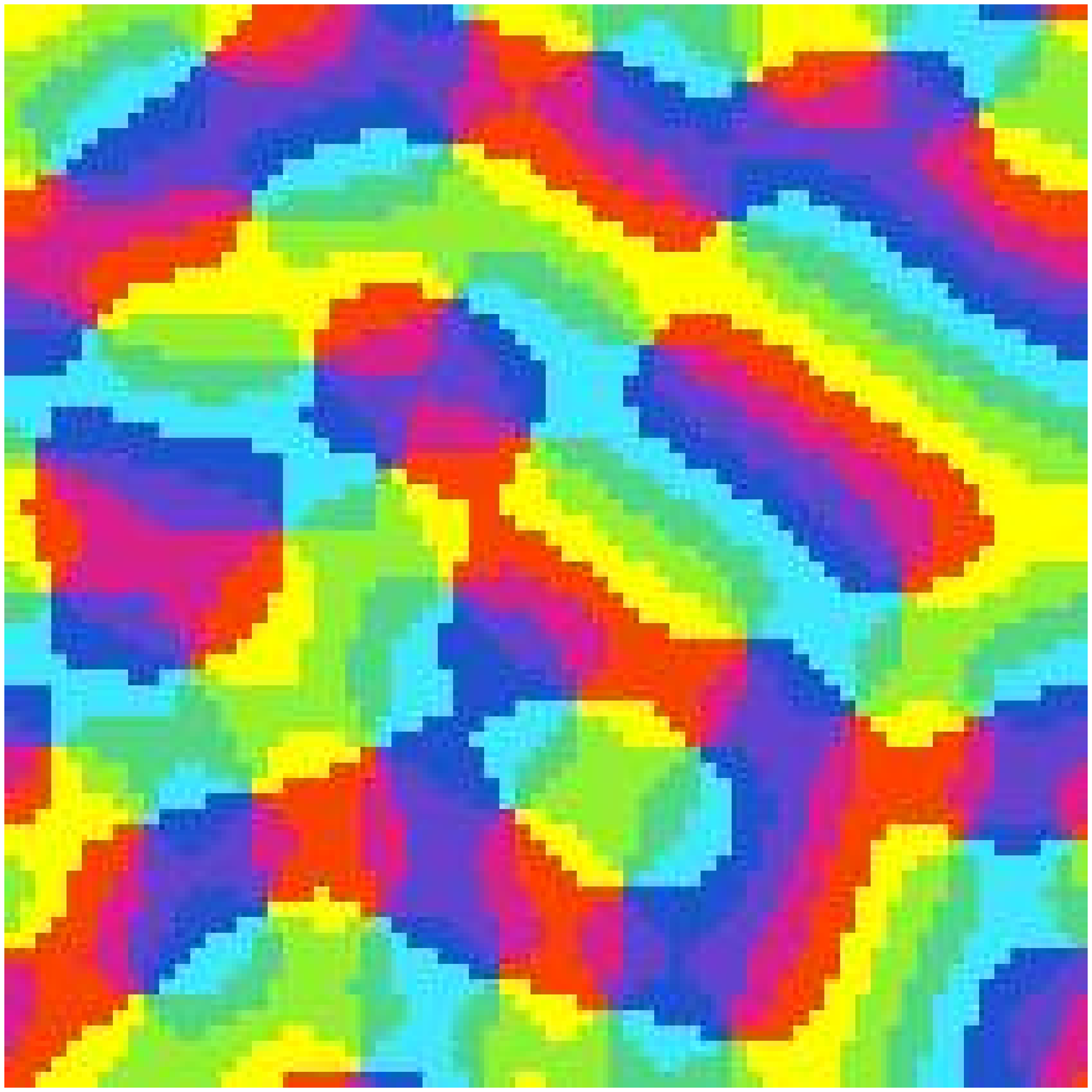} \end{minipage}
\\ \ \\
\begin{minipage}[b]{2cm} \centering (c) $k = 1.0$ \ \\ \ \\ \ \\ \end{minipage}
\begin{minipage}[b]{2cm} \includegraphics[width=1.9cm]{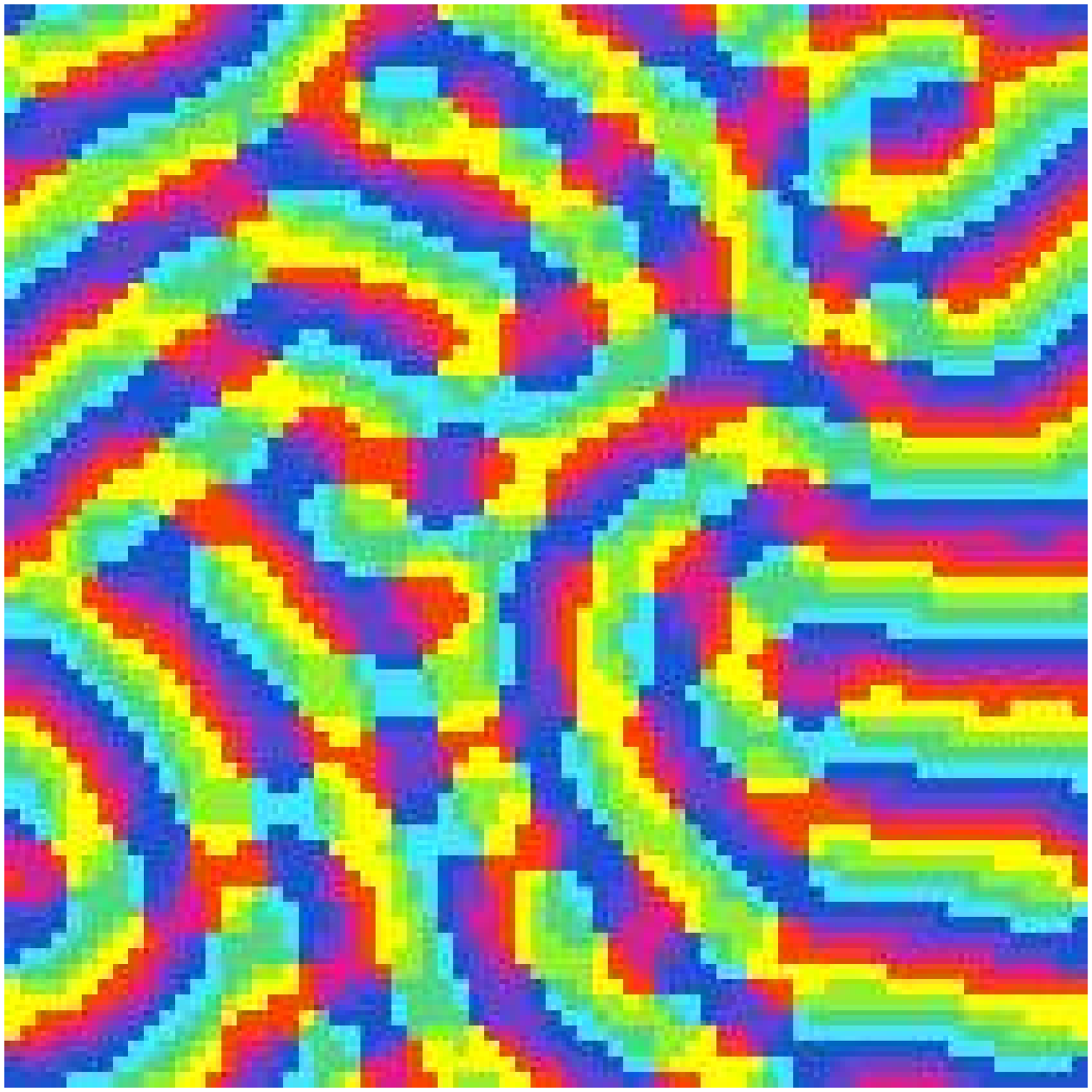} \end{minipage}
\begin{minipage}[b]{2cm} \includegraphics[width=1.9cm]{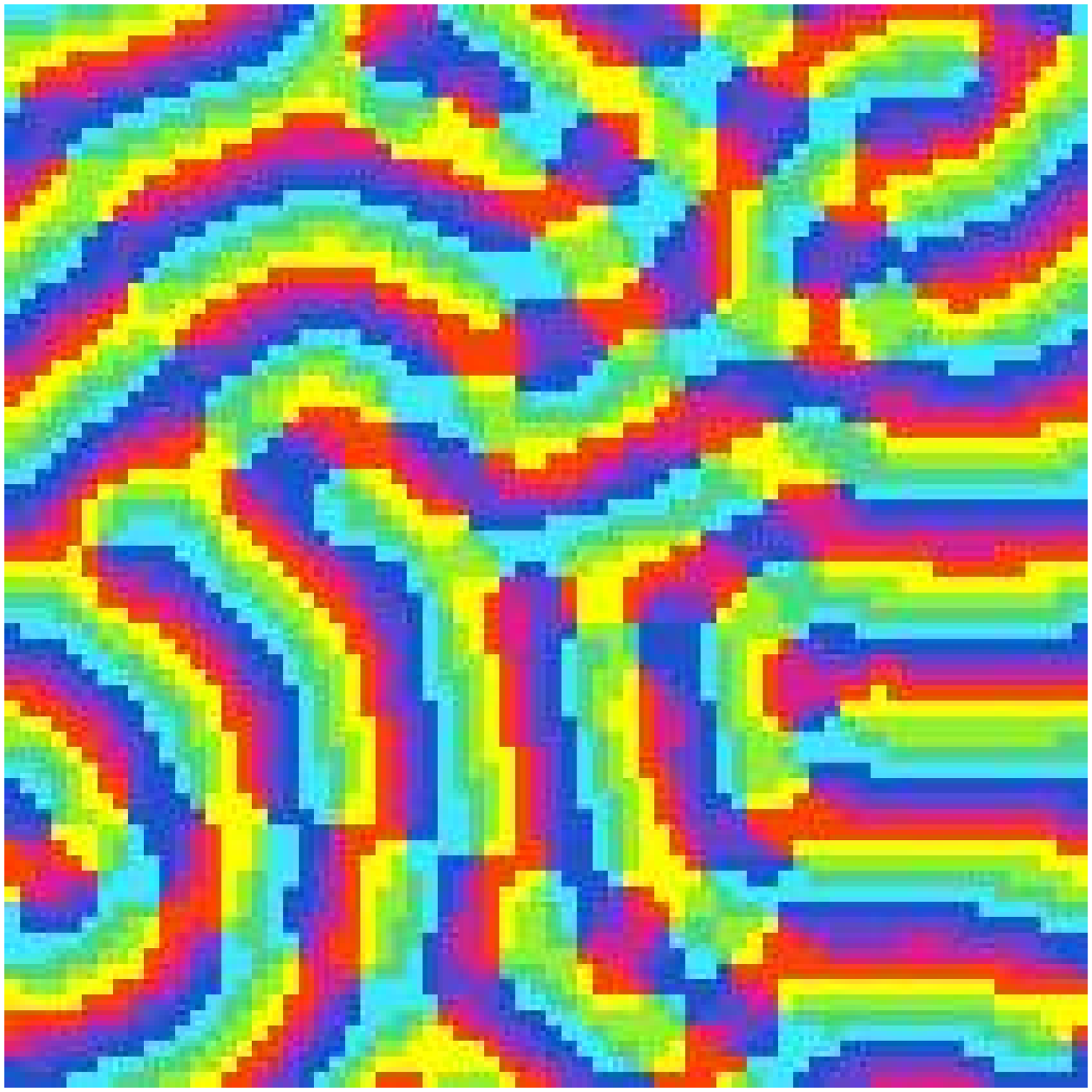} \end{minipage}
\begin{minipage}[b]{2cm} \includegraphics[width=1.9cm]{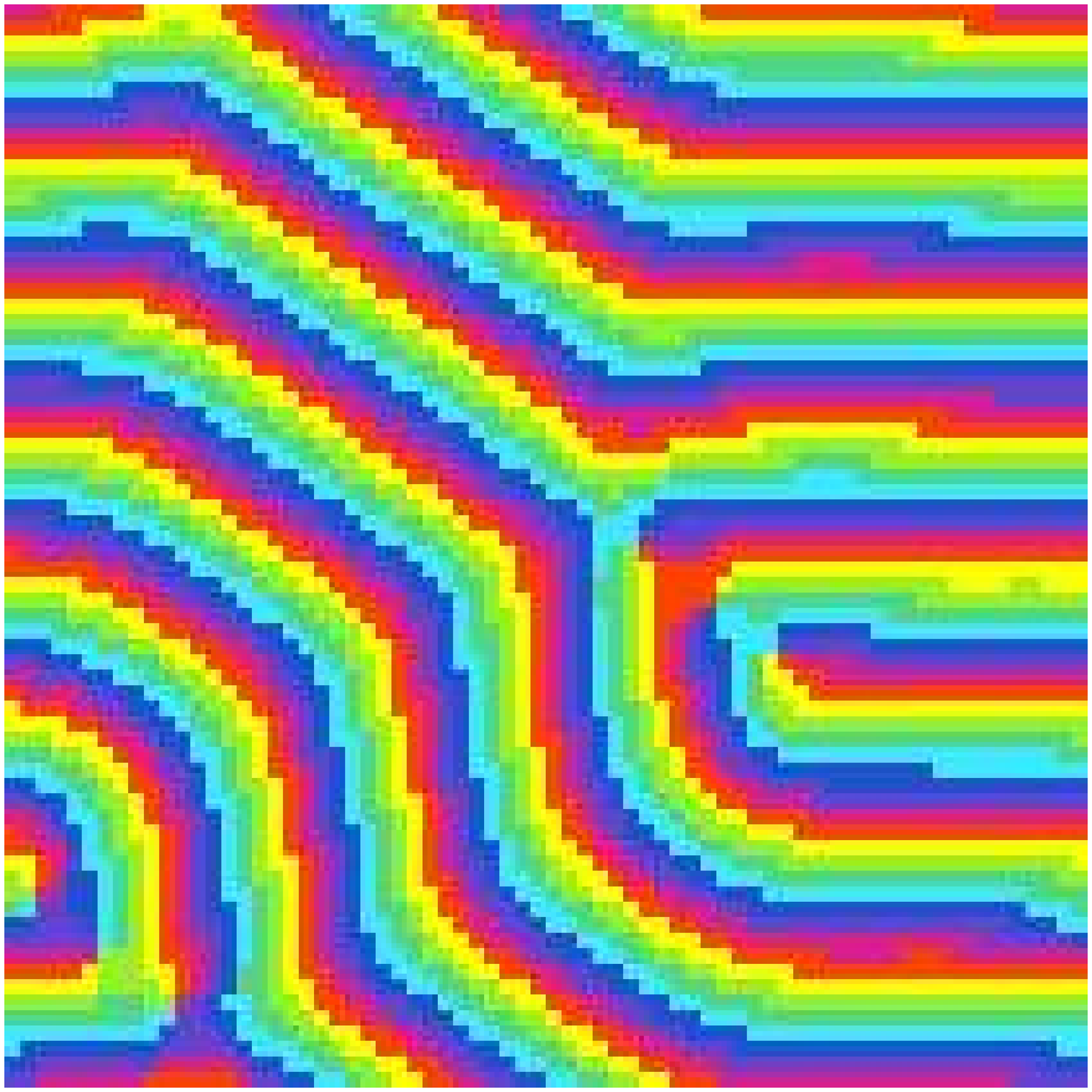} \end{minipage}
\\
\begin{minipage}[b]{2cm} \ \end{minipage}
\includegraphics[width=4cm]{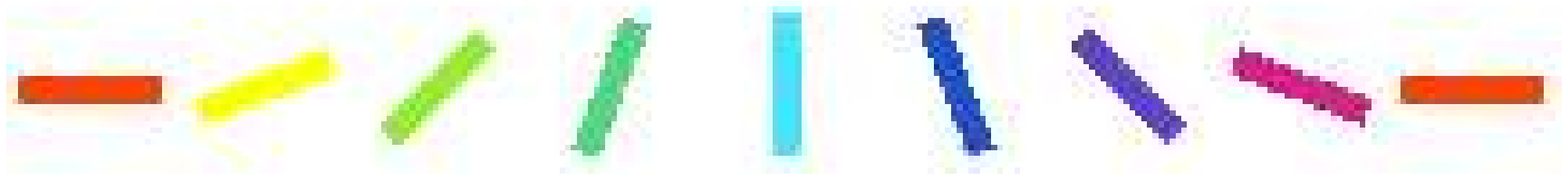}
\caption{\label{fig:op} Simulation results of the OP map using
  Eq.~(\ref{eq:Dynamics}).
  Maps are generated with $\sigma^2=6$, $\v=10^{-3}$, $\mu=0$, periodic
  boundary conditions and an initially random state in $70\times 70$ lattice.}
\end{figure}

The Hamiltonian in Eq.(\ref{eq:Hamiltonian}) can be easily diagonalized in the
momentum space:
\bn H=-\sum_\vq \tJ(\vq)\bS_\vq\cdot\bS_{-\vq}
  -\sum_\vq \bh_\vq\cdot \bS_{-\vq}, \en
where $\tJ(\vq)=\sum_\vr J(\vr)e^{-i\qcr}$,
$\bS_\vq=\f{1}{\sqrt{N}}\sum_i \bS_i e^{-i \qcr_i}$ and
$\bh_\vq=\f{1}{\sqrt{N}}\sum_i \bh_i e^{-i \qcr_i}$.
We assume that the LGN stimuli for a given neuron is uniform with no preferred
angle, so that the second term in Eq.~(\ref{eq:Hamiltonian}) is neglected
without random fluctuations effect.
In the continum limit, we obtain
\bn \tJ(\vq)\simeq \pi\v\f{\s^2}{a^2}(1-2k+k\s^2q^2)e^{-\s^2q^2/2}, 
  \label{eq:Jq} \en
where $a$ is the lattice constant.
$\tJ(\vq)$ has the maximum at $q^\ast=0$ for $k<k_c$ (=1/4) and at
$q^\ast=\f{1}{\s}\sqrt{4-1/k}$ for $k>k_c$ (Fig.~\ref{fig:Jq}).
Therefore, there is a threshold depending on $k_c$ below which OP slabs or OD
bands are forbidden.
Above $k_c$, columnar patterns emerge with the wavelength $\L=2\pi/q^\ast$,
which decreases as $k$ increases (Fig.~\ref{fig:op}~(b)~and~(c)). \par

In the simulations of the OP map formation starting from a random state by
Eq.~(\ref{eq:Dynamics}), dense pinwheels first emerge with the spin-waves for
$k<k_c$ (Fig.~\ref{fig:op}~(a)) as for the classical XY model.
Then pinwheels start to annihilate in pairs and eventually the map approaches
a homogeneous state.
For $k>k_c$ (Fig.~\ref{fig:op}~(b) and (c)), pinwheels emerge with the columnar
patterns and are annihilated in time as well.
The OP map will eventually approach the equilibrium state that is composed of
a uni-directional plane-wave, the winner in the competition among
$|\vq|=q^\ast$ states. \par

By using the parabolic behavior of $-\tJ(q)$ near the minimum point $q^\ast$,
the Hamiltonian can be approximated as
\bn H\simeq-\sum_{\vq}\left[\tJ(q^\ast)+\f{1}{2}\tJ''(q^\ast)
  |\vq-\vq^\ast|^2\right]\bS_{\vq}\cdot \bS_{-\vq} \en
or the effective Hamiltonian as a function of the preferred orentation $\phi$
is given by
\bn H[\phi]\simeq-NJ_s+\f{J_p}{2}\int d\vr|2\vN\phi-\vq^\ast|^2,
  \label{eq:app} \en
where $J_s=\tJ(q^\ast)$,
$J_p=-\tJ''(q^\ast)/a^2$ ($=2\pi\v\f{\s^4}{a^4}(4k-1)\exp(-2+1/2k)$ for
$k>k_c$), and both $J_s$ and $J_p$ are positive for all $k$.
The term $\vq^\ast$ in Eq.~(\ref{eq:app}), which makes plane-wave solutions,
does not contribute to the pinwheel formation energy since
$\bN\times\bN\phi_{pw}=0$ and the line integral around any contour vanishes by
Stoke's theorem.
Just adapting the results in vortex dynamics~\cite{Kosterlitz1973}, we can
obtain the change in free energy due to the formation of a pinwheel,
$\Delta G=(\pi J_p-2k_BT)\ln(L/a)$, and the phase transition temperature,
$T_{KT}=\pi J_p/2k_B$. \par

The visual cortex arises through activity-dependent refinement of initially
unselective patterns of synaptic connections, whereas dense pinwheels emerge
when orientation selectivity is first established and the density of pinwheels
decreases by annihilation in time.
The observed pinwheel densities differ in several species and such difference
in the pinwheel annihilation rates has been discussed by Wolf
{\em et al.}~\cite{Wolf1998}.
Now we can predict the relative pinwheel annihilation rates in terms of
$\Delta G$ or $T_{KT}$:
(i) As $k$ increases for $k>k_c$, $J_p$ increases and the pinwheels become more
unstable. 
For fixed $\v$ and $\s$, $\L$ decreases as $k$ increases and the map with a
narrower wavelength relaxes to the equilibrium state more rapidly
(Fig.~\ref{fig:op}~(b)~\&~(c)).
(ii) As $\s$ increases, pinwheels become more unstable.
But $\L$ is proportional to $\s$, and the map with a narrower wavelength
relaxes to the equilibrium state more slowly in this case.
(iii) The parameter $\v$ can be regardes the activity rate responding to
external stimuli or the learning rate due to the Hebbian rule.
The annihilation rate becomes larger for larger $\v$.
(iv) Thermal fluctuations may lead to the persistence of the pinwheel structure
but they also disturb the map organization.
(v)
% The interaction energy of a pair of pinwheel-antipinwheel is
%$E_{pair}(\vr_1,\vr_2)=-2\pi J_p\ln(|\vr_1-\vr_2|/a)$.
%But there is a pinwheel annihilation mechanism by collisions not only between
There are pinwheel annihilation mechanisms by collisions not only between
opposite vortices but also with area boundaries.
The probability of a collision with area boundaries decreases as the lattice
size increases for randomly moving pinwheels.
(vi) To include the interactions between OD and OP columns, our model has to be
extended to the $O(3)$ symmetry or Heisenberg model.
The classical anisotropic Heisenberg model is described by
\bn H=-K\sum_{\lg ij\rg}(S_i^xS_j^x+S_i^yS_j^y+\lambda S_i^zS_j^z), \en
where $K>0$.
For this model, $T_{KT}$ is known to approach 0 as $\lambda$ approaches
$1$~\cite{Hikami1980}.
This result can be translated as follows: the pinwheel structures last longer
or even become stabilized through strong activity of OD column. \par

\begin{figure}[b]
\begin{minipage}[b]{2.5cm} \includegraphics[width=2.5cm]{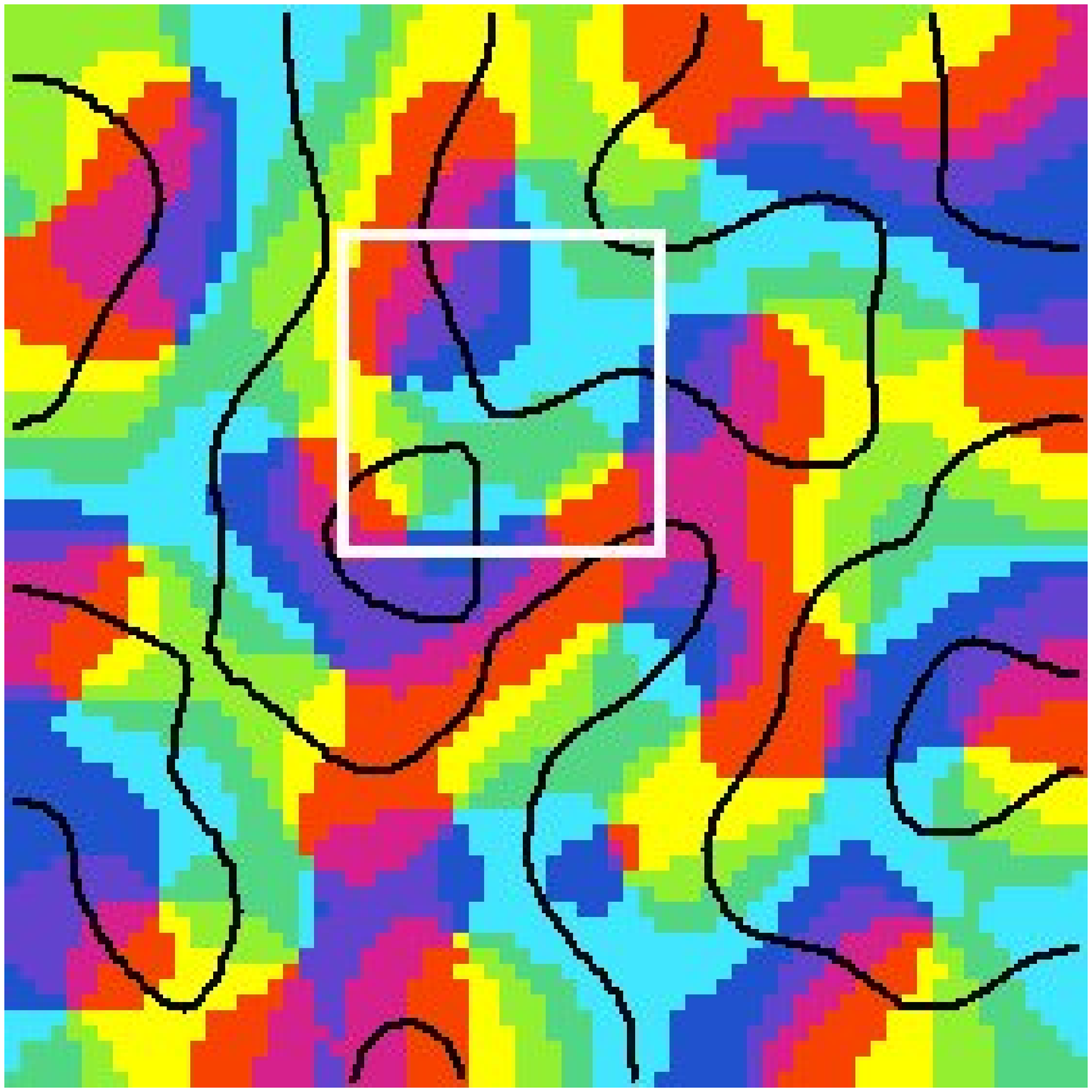} (a) \end{minipage}
\ \ \ 
\begin{minipage}[b]{2.5cm} \includegraphics[width=2.5cm]{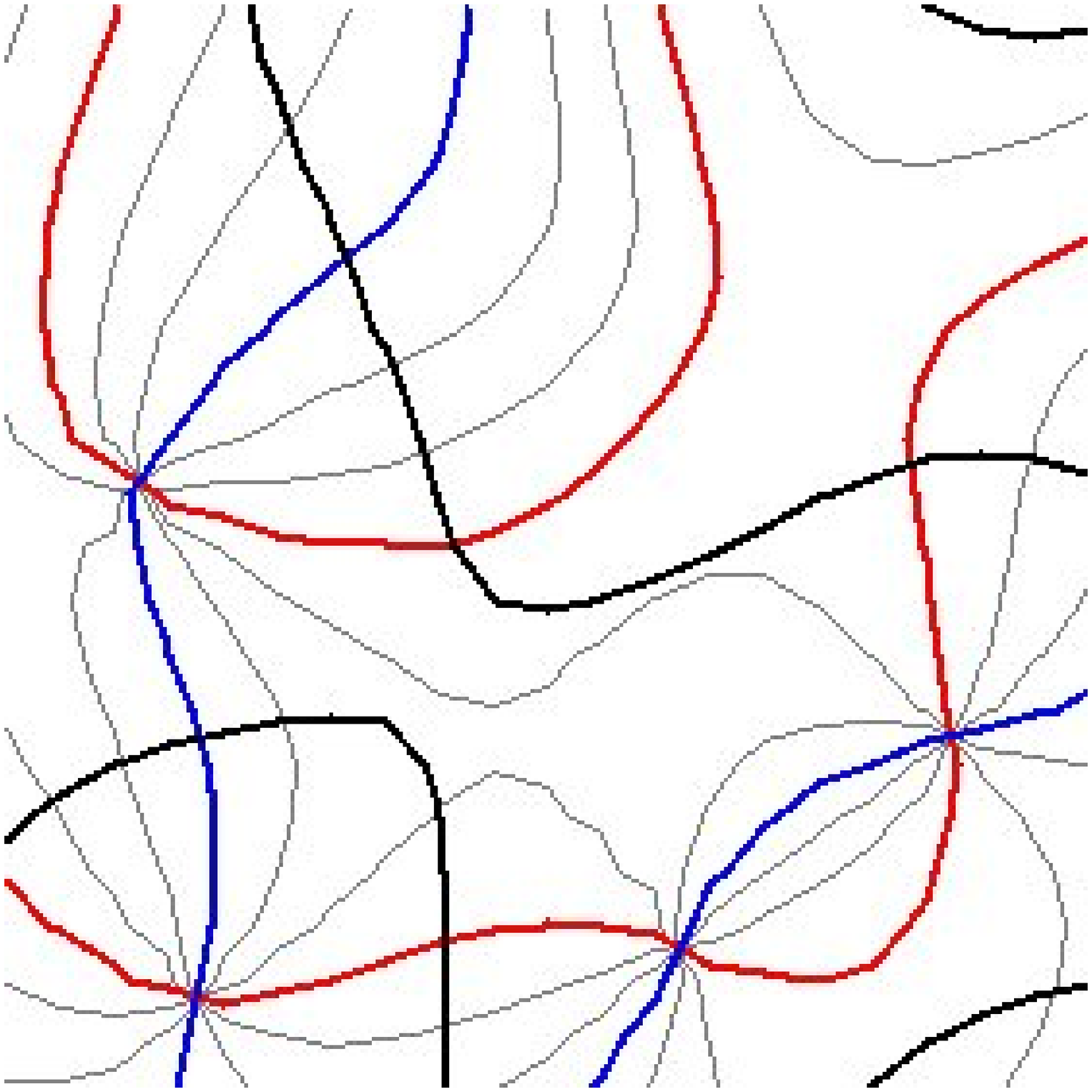} (b) \end{minipage}
\caption{\label{fig:o3}
  (a) An simulation result of OP and OD columnar patterns using
  Eq.~(\ref{eq:Hamiltonian}) with $\bS=(S_x,S_y,S_z)$ and (b) its detailed
  contour feature in a white rectangle (non-periodic boundary conditions).
  The blue lines correspond to $S_x=0$ ($\phi=\f{\pi}{4}$ or $\f{3\pi}{4}$)
  domains and the red lines, $S_y=0$ ($\phi=0$ or $\f{\pi}{2}$) domains.
  The black lines are the borders of opposite OD (or $S_z=0$) domains in both
  figures.
}
\end{figure}

The extension to the Heisenberg model also explains the observed orthogonal
property between the borders of OD bands and the IOCs.
Let us consider the gradient or normal vectors of IOCs at $S_x=0$ and $S_y=0$.
These two vectors intersect perpendicularly at pinwheels.
The borders of the opposite OD domains can be represented as $S_z=0$, which
will meet also perpendicularly with other contours, $S_x=0$ or $S_y=0$.
Therefore, the borders of OD bands are mathematically equivalent to IOCs and
intersect perpendicularly with each other (Fig.~\ref{fig:o3}). \par

\begin{figure}[t]
\begin{minipage}[b]{1cm} $g(r)$ \ \\ \ \\ \ \\ \ \\ \ \\ \ \\ \ \\ \end{minipage}
\begin{minipage}[b]{5cm} \includegraphics[width=5cm]{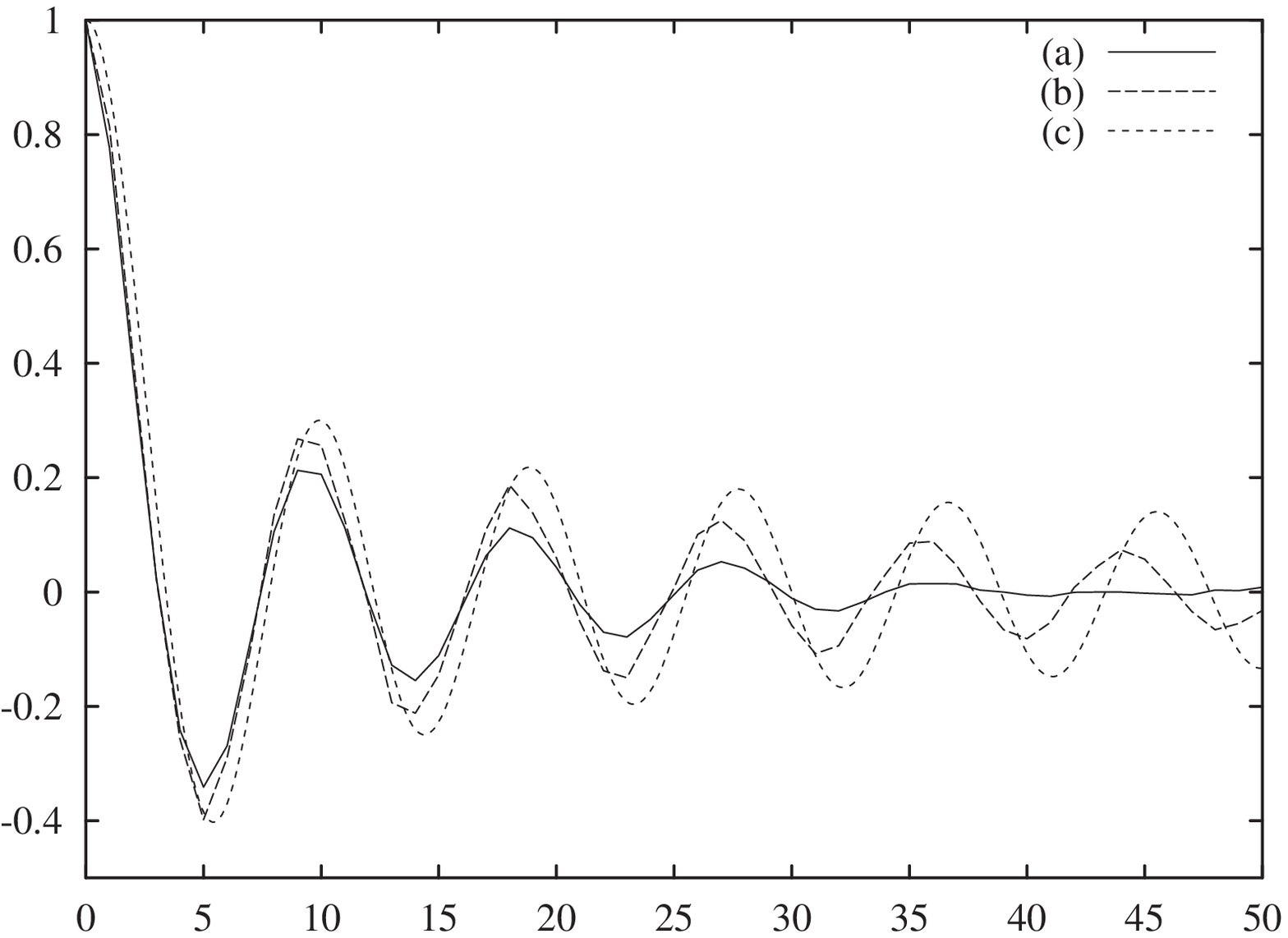} $r$ \end{minipage}
\caption{ \label{fig:corr} (a) Normalized autocorrelation function $g(r)$ of
  the simulation results in Fig.~\ref{fig:op}~(c) at $t=5$ and (b) $t=100$.
  (c) The zeroth Bessel function $J_0(q^\ast r)$. }
\ \\
\includegraphics[width=6cm]{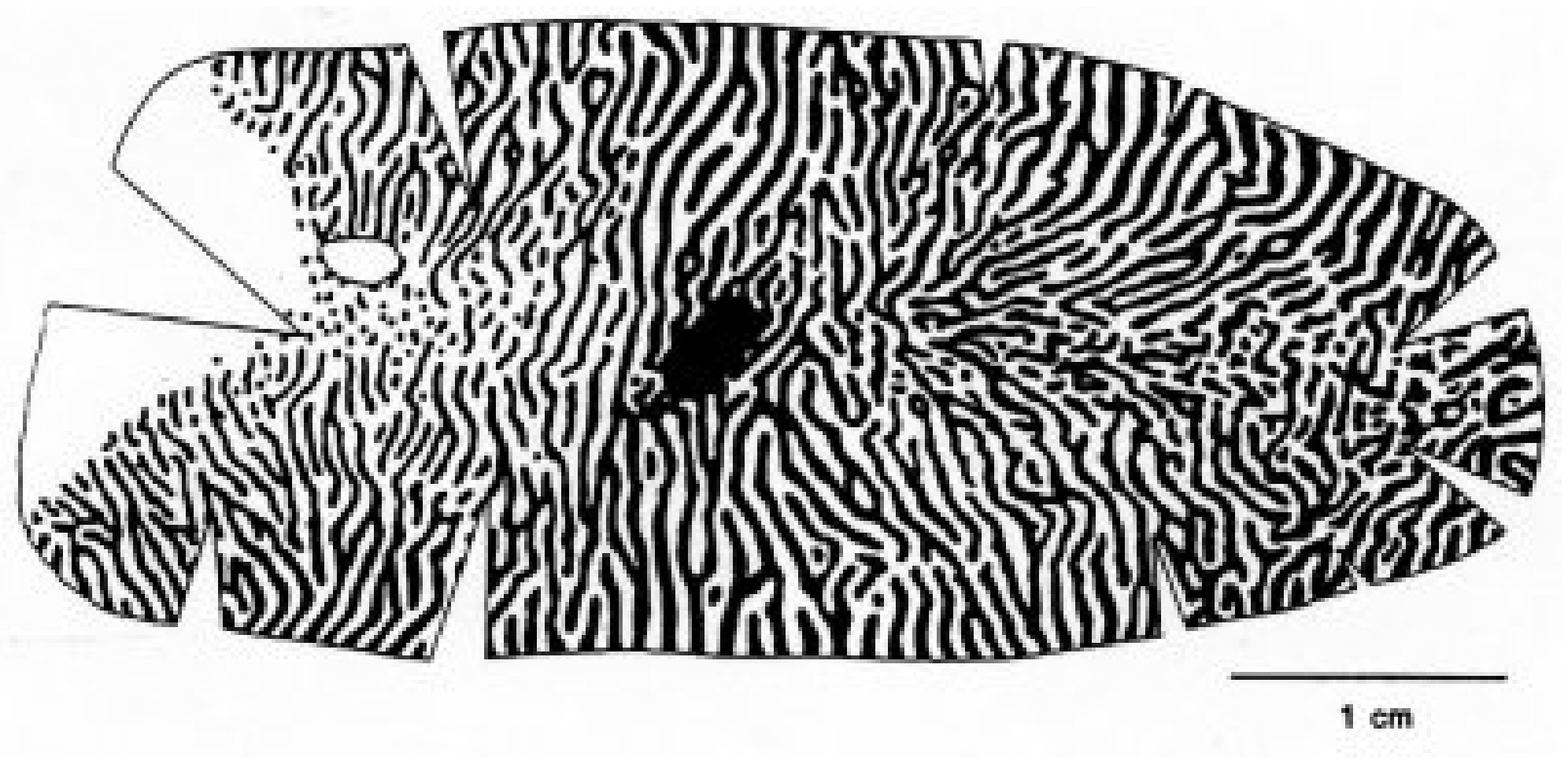}
\caption{\label{fig:macaque} The complete pattern of OD stripes in the striate
  cortex of a macaque monkey. There is a strong tendency for the stripes to
  meet the margin of striate cortex at steep or right angles
  (LeVay~\cite{LeVay1985}, Copyright 1985 by the Society for Neuroscience). }
\end{figure}

The correlation function in the OP maps is obtained from Eq.~(\ref{eq:app}) as
\bn \lg\bS(r)\cdot\bS(0)\rg=g(r)\simeq J_0(q^\ast r)\
  \left(\f{r}{a}\right)^{-k_BT/2\pi J_p}, \label{eq:corr} \en
%  \left(\f{r}{a}\right)^{-\eta(T)}, \label{eq:corr} \en
%where $\eta(T)=k_BT/2\pi J_p$ and $J_0$ the zeroth Bessel function that is the
where $J_0$ is the zeroth Bessel function.
As the map relaxes to the equilibrium state, the correlation function is
expected to approach the distribution in Eq.~(\ref{eq:corr}).
This is consistent with the experimens~\cite{Erwin1995} and the numerical
simulations (Fig.~\ref{fig:corr}).\par

The perpendicularity with the margin of the striate cortex is derived from the
equilibrium condition $\d H/\d\phi\sim 0$ or $\nabla^2\phi\sim 0$.
The normal component of $\bN\phi$ vanishes at the area boundary since the
integral along a narrow rectangular loop over the area boundary,
$\oint_C\bN\phi\cdot d\vec{n}$, vanishes due to the divergence theorem.
Such perpendicularity with the area boundary is also manifested in other static
field solutions, such as the magnetic field, and is consistent with the
observed OD patterns {\em in vivo} (Fig.~\ref{fig:macaque}). \par

The interactions in cortical circuitry and the synaptic plasticity are more
complex processes {\em in vivo} and various cortical modification rules are
postulated by other models.
Competitive Hebbian models also provide somewhat successful description of
the map formation based on the different assumption that neurons compete with
neighborhoods and get more stimuli close to their preferred angle.
Then the second term in Eq.(\ref{eq:Hamiltonian}) from LGN stimuli will no
longer vanish upon averaging.
However, both non-competitive and competitive models can generate and predict
similar pattern properties without much differences.
The FBM model, named from the fibre bundle theory in manifold, explains the
emerging patterns in cortical map with differential geometric concept that the
major features of self-organizing map are determined not by the detailed
interactions of processing elements but by lattice geometry, symmetry group and
connections in phase space (that is $\vq^\ast$ in our problem). \par

The correlations between the OP and OD columns through $O(3)$ symmetry come
from the normalization of synapse strength.
If the OP and OD columns have different interaction strength, the total system
would be described by an easy-plane (or XY symmetry) Heisenberg model.
The comprehensive understanding of distance dependent anisotropic Heisenberg
models is not an easy by itself and is as an important issue also in condense
matter physics, the study of which will have important implications in
understanding the effects of the interactions between OP and OD columns in the
visual cortex.

We are grateful for discussions with Professor S. Tanaka.
This work was supported by the Ministry of Science and Technology and the
Ministry of Education.

\bibliography{letter}

\begin{thebibliography}{28}
\expandafter\ifx\csname natexlab\endcsname\relax\def\natexlab#1{#1}\fi
\expandafter\ifx\csname bibnamefont\endcsname\relax
  \def\bibnamefont#1{#1}\fi
\expandafter\ifx\csname bibfnamefont\endcsname\relax
  \def\bibfnamefont#1{#1}\fi
\expandafter\ifx\csname citenamefont\endcsname\relax
  \def\citenamefont#1{#1}\fi
\expandafter\ifx\csname url\endcsname\relax
  \def\url#1{\texttt{#1}}\fi
\expandafter\ifx\csname urlprefix\endcsname\relax\def\urlprefix{URL }\fi
\providecommand{\bibinfo}[2]{#2}
\providecommand{\eprint}[2][]{\url{#2}}

\bibitem[{\citenamefont{Hubel and Wiesel}(1977)}]{Hubel1977}
\bibinfo{author}{\bibfnamefont{D.}~\bibnamefont{Hubel}} \bibnamefont{and}
  \bibinfo{author}{\bibfnamefont{T.~N.} \bibnamefont{Wiesel}},
  \bibinfo{journal}{Proc.\ Roy.\ Soc.~(London) B}
  \textbf{\bibinfo{volume}{278}}, \bibinfo{pages}{377} (\bibinfo{year}{1977}).

\bibitem[{\citenamefont{LeVay et~al.}(1978)\citenamefont{LeVay, Stryker, and
  Shatz}}]{LeVay1978}
\bibinfo{author}{\bibfnamefont{S.}~\bibnamefont{LeVay}},
  \bibinfo{author}{\bibfnamefont{M.}~\bibnamefont{Stryker}}, \bibnamefont{and}
  \bibinfo{author}{\bibfnamefont{C.}~\bibnamefont{Shatz}},
  \bibinfo{journal}{J.~Comp.\ Neurol.} \textbf{\bibinfo{volume}{179}},
  \bibinfo{pages}{223} (\bibinfo{year}{1978}).

\bibitem[{\citenamefont{Stryker et~al.}(1978)\citenamefont{Stryker, Sherk,
  Leventhal, and Hirsch}}]{Stryker1978}
\bibinfo{author}{\bibfnamefont{M.~P.} \bibnamefont{Stryker}},
  \bibinfo{author}{\bibfnamefont{H.}~\bibnamefont{Sherk}},
  \bibinfo{author}{\bibfnamefont{A.~G.} \bibnamefont{Leventhal}},
  \bibnamefont{and} \bibinfo{author}{\bibfnamefont{H.~V.~B.}
  \bibnamefont{Hirsch}}, \bibinfo{journal}{J.\ Neurophysiol.}
  \textbf{\bibinfo{volume}{41}}, \bibinfo{pages}{896} (\bibinfo{year}{1978}).

\bibitem[{\citenamefont{Blasdel}(1992)}]{Blasdel1992}
\bibinfo{author}{\bibfnamefont{G.~G.} \bibnamefont{Blasdel}},
  \bibinfo{journal}{J.\ Neurosci.} \textbf{\bibinfo{volume}{12}},
  \bibinfo{pages}{3115} (\bibinfo{year}{1992}).

\bibitem[{\citenamefont{Blasdel and Salama}(1986)}]{Blasdel1986}
\bibinfo{author}{\bibfnamefont{G.~G.} \bibnamefont{Blasdel}} \bibnamefont{and}
  \bibinfo{author}{\bibfnamefont{G.}~\bibnamefont{Salama}},
  \bibinfo{journal}{Nature} \textbf{\bibinfo{volume}{321}},
  \bibinfo{pages}{579} (\bibinfo{year}{1986}).

\bibitem[{\citenamefont{Grinvald et~al.}(1986)\citenamefont{Grinvald, Lieke,
  Frostig, Gilbert, and Wiesel}}]{Grinvald1986}
\bibinfo{author}{\bibfnamefont{A.}~\bibnamefont{Grinvald}},
  \bibinfo{author}{\bibfnamefont{E.}~\bibnamefont{Lieke}},
  \bibinfo{author}{\bibfnamefont{R.~P.} \bibnamefont{Frostig}},
  \bibinfo{author}{\bibfnamefont{C.}~\bibnamefont{Gilbert}}, \bibnamefont{and}
  \bibinfo{author}{\bibfnamefont{T.}~\bibnamefont{Wiesel}},
  \bibinfo{journal}{Nature} \textbf{\bibinfo{volume}{324}},
  \bibinfo{pages}{361} (\bibinfo{year}{1986}).

\bibitem[{\citenamefont{Erwin et~al.}(1995)\citenamefont{Erwin, Obermayer, and
  Schulten}}]{Erwin1995}
\bibinfo{author}{\bibfnamefont{E.}~\bibnamefont{Erwin}},
  \bibinfo{author}{\bibfnamefont{K.}~\bibnamefont{Obermayer}},
  \bibnamefont{and} \bibinfo{author}{\bibfnamefont{K.}~\bibnamefont{Schulten}},
  \bibinfo{journal}{Neural\ comput.} \textbf{\bibinfo{volume}{7}},
  \bibinfo{pages}{425} (\bibinfo{year}{1995}).

\bibitem[{\citenamefont{Swindale}(1996)}]{Swindale1996}
\bibinfo{author}{\bibfnamefont{N.~V.} \bibnamefont{Swindale}},
  \bibinfo{journal}{Network:\ Comput.\ Neural\ Syst.}
  \textbf{\bibinfo{volume}{7}}, \bibinfo{pages}{161} (\bibinfo{year}{1996}).

\bibitem[{\citenamefont{Swindale et~al.}(1987)\citenamefont{Swindale,
  Matsubara, and Cynader}}]{Swindale1987}
\bibinfo{author}{\bibfnamefont{N.~V.} \bibnamefont{Swindale}},
  \bibinfo{author}{\bibfnamefont{J.~A.} \bibnamefont{Matsubara}},
  \bibnamefont{and} \bibinfo{author}{\bibfnamefont{M.~S.}
  \bibnamefont{Cynader}}, \bibinfo{journal}{J.~Neurosci.}
  \textbf{\bibinfo{volume}{7}}, \bibinfo{pages}{1414} (\bibinfo{year}{1987}).

\bibitem[{\citenamefont{Bonhoeffer and Grinvald}(1991)}]{Bonhoeffer1991}
\bibinfo{author}{\bibfnamefont{T.}~\bibnamefont{Bonhoeffer}} \bibnamefont{and}
  \bibinfo{author}{\bibfnamefont{A.}~\bibnamefont{Grinvald}},
  \bibinfo{journal}{Nature} \textbf{\bibinfo{volume}{353}},
  \bibinfo{pages}{429} (\bibinfo{year}{1991}).

\bibitem[{\citenamefont{Maldonado et~al.}(1997)\citenamefont{Maldonado,
  G{\"o}decke, Gray, and Bonhoeffer}}]{Maldonado1997}
\bibinfo{author}{\bibfnamefont{P.~E.} \bibnamefont{Maldonado}},
  \bibinfo{author}{\bibfnamefont{I.}~\bibnamefont{G{\"o}decke}},
  \bibinfo{author}{\bibfnamefont{C.~M.} \bibnamefont{Gray}}, \bibnamefont{and}
  \bibinfo{author}{\bibfnamefont{T.}~\bibnamefont{Bonhoeffer}},
  \bibinfo{journal}{Science} \textbf{\bibinfo{volume}{276}},
  \bibinfo{pages}{1551} (\bibinfo{year}{1997}).

\bibitem[{\citenamefont{Obermayer et~al.}(1992)\citenamefont{Obermayer,
  Blasdel, and Schulten}}]{Obermayer1992}
\bibinfo{author}{\bibfnamefont{K.}~\bibnamefont{Obermayer}},
  \bibinfo{author}{\bibfnamefont{G.~G.} \bibnamefont{Blasdel}},
  \bibnamefont{and} \bibinfo{author}{\bibfnamefont{K.}~\bibnamefont{Schulten}},
  \bibinfo{journal}{Phys.\ Rev.\ A} \textbf{\bibinfo{volume}{45}},
  \bibinfo{pages}{7568} (\bibinfo{year}{1992}).

\bibitem[{\citenamefont{Durbin and Mitchison}(1990)}]{Durbin1990}
\bibinfo{author}{\bibfnamefont{R.}~\bibnamefont{Durbin}} \bibnamefont{and}
  \bibinfo{author}{\bibfnamefont{G.}~\bibnamefont{Mitchison}},
  \bibinfo{journal}{Nature} \textbf{\bibinfo{volume}{343}},
  \bibinfo{pages}{341} (\bibinfo{year}{1990}).

\bibitem[{\citenamefont{Scherf et~al.}(1999)\citenamefont{Scherf, Pawelzik,
  Wolf, and Geisel}}]{Scherf1999}
\bibinfo{author}{\bibfnamefont{O.}~\bibnamefont{Scherf}},
  \bibinfo{author}{\bibfnamefont{K.}~\bibnamefont{Pawelzik}},
  \bibinfo{author}{\bibfnamefont{F.}~\bibnamefont{Wolf}}, \bibnamefont{and}
  \bibinfo{author}{\bibfnamefont{T.}~\bibnamefont{Geisel}},
  \bibinfo{journal}{Phys.\ Rev.\ E} \textbf{\bibinfo{volume}{59}},
  \bibinfo{pages}{6977} (\bibinfo{year}{1999}).

\bibitem[{\citenamefont{Goodhill and Cimponeriu}(2000)}]{Goodhill2000}
\bibinfo{author}{\bibfnamefont{G.~J.} \bibnamefont{Goodhill}} \bibnamefont{and}
  \bibinfo{author}{\bibfnamefont{A.}~\bibnamefont{Cimponeriu}},
  \bibinfo{journal}{Network:\ Comput.\ Neural\ Syst.}
  \textbf{\bibinfo{volume}{11}}, \bibinfo{pages}{153} (\bibinfo{year}{2000}).

\bibitem[{\citenamefont{Hoffs{\"u}mmer
  et~al.}(1995)\citenamefont{Hoffs{\"u}mmer, Wolf, and
  Geisel}}]{Hoffsummer1995}
\bibinfo{author}{\bibfnamefont{F.}~\bibnamefont{Hoffs{\"u}mmer}},
  \bibinfo{author}{\bibfnamefont{F.}~\bibnamefont{Wolf}}, \bibnamefont{and}
  \bibinfo{author}{\bibfnamefont{T.}~\bibnamefont{Geisel}},
  \bibinfo{journal}{Neurol.\ Conf.} \textbf{\bibinfo{volume}{1}},
  \bibinfo{pages}{97} (\bibinfo{year}{1995}).

\bibitem[{\citenamefont{Wolf and Geisel}(1998)}]{Wolf1998}
\bibinfo{author}{\bibfnamefont{F.}~\bibnamefont{Wolf}} \bibnamefont{and}
  \bibinfo{author}{\bibfnamefont{T.}~\bibnamefont{Geisel}},
  \bibinfo{journal}{Nature} \textbf{\bibinfo{volume}{395}}, \bibinfo{pages}{73}
  (\bibinfo{year}{1998}).

\bibitem[{\citenamefont{Daw}(1998)}]{Daw1998}
\bibinfo{author}{\bibfnamefont{N.~W.} \bibnamefont{Daw}},
  \bibinfo{journal}{Nature} \textbf{\bibinfo{volume}{395}}, \bibinfo{pages}{20}
  (\bibinfo{year}{1998}).

\bibitem[{\citenamefont{Obermayer and Blasdel}(1993)}]{Obermayer1993}
\bibinfo{author}{\bibfnamefont{K.}~\bibnamefont{Obermayer}} \bibnamefont{and}
  \bibinfo{author}{\bibfnamefont{G.~G.} \bibnamefont{Blasdel}},
  \bibinfo{journal}{J.\ Neurosci.} \textbf{\bibinfo{volume}{13}},
  \bibinfo{pages}{4114} (\bibinfo{year}{1993}).

\bibitem[{\citenamefont{LeVay et~al.}(1985)\citenamefont{LeVay, Connolly,
  Houde, and Essen}}]{LeVay1985}
\bibinfo{author}{\bibfnamefont{S.}~\bibnamefont{LeVay}},
  \bibinfo{author}{\bibfnamefont{D.~H.} \bibnamefont{Connolly}},
  \bibinfo{author}{\bibfnamefont{J.}~\bibnamefont{Houde}}, \bibnamefont{and}
  \bibinfo{author}{\bibfnamefont{D.~C.~V.} \bibnamefont{Essen}},
  \bibinfo{journal}{J.\ Neurosci.} \textbf{\bibinfo{volume}{5}},
  \bibinfo{pages}{486} (\bibinfo{year}{1985}).

\bibitem[{\citenamefont{Kosterlitz and Thouless}(1973)}]{Kosterlitz1973}
\bibinfo{author}{\bibfnamefont{J.~M.} \bibnamefont{Kosterlitz}}
  \bibnamefont{and} \bibinfo{author}{\bibfnamefont{D.~J.}
  \bibnamefont{Thouless}}, \bibinfo{journal}{J.\ Phys.\ C}
  \textbf{\bibinfo{volume}{6}}, \bibinfo{pages}{1181} (\bibinfo{year}{1973}).

\bibitem[{\citenamefont{Calvin}(1998)}]{Calvin1998}
\bibinfo{author}{\bibfnamefont{W.~H.} \bibnamefont{Calvin}}, in
  \emph{\bibinfo{booktitle}{The handbook of brain theory and neural networks}},
  edited by \bibinfo{editor}{\bibfnamefont{M.~A.} \bibnamefont{Arbib}}
  (\bibinfo{publisher}{MIT Press}, \bibinfo{year}{1998}), pp.
  \bibinfo{pages}{269--272}.

\bibitem[{\citenamefont{Lund et~al.}(1998)\citenamefont{Lund, Wu, and
  Levitt}}]{Lund1998}
\bibinfo{author}{\bibfnamefont{J.~S.} \bibnamefont{Lund}},
  \bibinfo{author}{\bibfnamefont{Q.}~\bibnamefont{Wu}}, \bibnamefont{and}
  \bibinfo{author}{\bibfnamefont{J.~B.} \bibnamefont{Levitt}}, in
  \emph{\bibinfo{booktitle}{The handbook of brain theory and neural networks}},
  edited by \bibinfo{editor}{\bibfnamefont{M.~A.} \bibnamefont{Arbib}}
  (\bibinfo{publisher}{MIT Press}, \bibinfo{year}{1998}), pp.
  \bibinfo{pages}{1016--1021}.

\bibitem[{\citenamefont{Monroe et~al.}(1990)\citenamefont{Monroe, Lucente, and
  Hourlland}}]{Monroe1990}
\bibinfo{author}{\bibfnamefont{J.~L.} \bibnamefont{Monroe}},
  \bibinfo{author}{\bibfnamefont{R.}~\bibnamefont{Lucente}}, \bibnamefont{and}
  \bibinfo{author}{\bibfnamefont{J.~P.} \bibnamefont{Hourlland}},
  \bibinfo{journal}{J.\ Phys.\ A} \textbf{\bibinfo{volume}{23}},
  \bibinfo{pages}{2555} (\bibinfo{year}{1990}).

\bibitem[{\citenamefont{Krech and Luijten}(2000)}]{Krech2000}
\bibinfo{author}{\bibfnamefont{M.}~\bibnamefont{Krech}} \bibnamefont{and}
  \bibinfo{author}{\bibfnamefont{E.}~\bibnamefont{Luijten}},
  \bibinfo{journal}{Phys.\ Rev.\ E} \textbf{\bibinfo{volume}{61}},
  \bibinfo{pages}{2058} (\bibinfo{year}{2000}).

\bibitem[{\citenamefont{Ifti et~al.}(2001)\citenamefont{Ifti, Li, Soukoulis,
  and Velgakis}}]{Ifti2001}
\bibinfo{author}{\bibfnamefont{M.}~\bibnamefont{Ifti}},
  \bibinfo{author}{\bibfnamefont{Q.}~\bibnamefont{Li}},
  \bibinfo{author}{\bibfnamefont{C.~M.} \bibnamefont{Soukoulis}},
  \bibnamefont{and} \bibinfo{author}{\bibfnamefont{M.~J.}
  \bibnamefont{Velgakis}}, \bibinfo{journal}{Mod.\ Phys.\ Lett.\ B}
  \textbf{\bibinfo{volume}{15}}, \bibinfo{pages}{895} (\bibinfo{year}{2001}).

\bibitem[{\citenamefont{Luijten and Bl{\"o}te}(1997)}]{Luijten1997}
\bibinfo{author}{\bibfnamefont{E.}~\bibnamefont{Luijten}} \bibnamefont{and}
  \bibinfo{author}{\bibfnamefont{H.~W.~J.} \bibnamefont{Bl{\"o}te}},
  \bibinfo{journal}{Phys.\ Rev.\ B} \textbf{\bibinfo{volume}{56}},
  \bibinfo{pages}{8545} (\bibinfo{year}{1997}).

\bibitem[{\citenamefont{Hikami and Tsuneto}(1980)}]{Hikami1980}
\bibinfo{author}{\bibfnamefont{S.}~\bibnamefont{Hikami}} \bibnamefont{and}
  \bibinfo{author}{\bibfnamefont{T.}~\bibnamefont{Tsuneto}},
  \bibinfo{journal}{Prog.\ Theor.\ Phys.} \textbf{\bibinfo{volume}{63}},
  \bibinfo{pages}{387} (\bibinfo{year}{1980}).

\end{thebibliography}

\end{document}